\documentclass[twocolumn,showpacs,preprintnumbers,amsmath,amssymb]{revtex4}

\usepackage{graphicx}
\usepackage{dcolumn}
\usepackage{bm}

\begin{document}
\preprint{APS/123-QED}
\title{Huge First-Order Metamagnetic Transition in the Paramagnetic Heavy-Fermion System CeTiGe}
\author{M. Deppe$^{1}$, S. Lausberg$^{1}$, F. Weickert$^{1,3}$, M. Brando$^{1}$, Y. Skourski$^{2}$, N. Caroca-Canales$^{1}$, C. Geibel$^{1}$, and F. Steglich$^{1}$}
\affiliation{$^{1}$ Max-Planck-Institute for Chemical Physics of Solids, D-01187 Dresden, Germany\\
$^{2}$ Dresden High Magnetic Field Laboratory (HLD), D-01328 Dresden, Germany\\
$^{3}$ MPA-CMMS, Los Alamos National Laboratory (LANL), Los Alamos, New Mexico 87545, USA}
\date{\today}
\begin{abstract}
We report on the observation of large, step-like anomalies in the magnetization ($\Delta M = 0.74$\,$\mu_{\rm B}$/Ce),
in the magnetostriction ($\Delta l/l_{0} = 2.0 \cdot 10^{-3}$), and in the magnetoresistance in polycrystals of the
paramagnetic heavy-fermion system CeTiGe at a critical magnetic field $\mu_0 H_c \approx $ 12.5\,T at low temperatures.
The size of these anomalies is much larger than those reported for the prototypical heavy-fermion metamagnet
CeRu$_2$Si$_2$. Furthermore, hysteresis between increasing and decreasing field data indicate a real thermodynamic,
first-order type of phase transition, in contrast to the crossover reported for CeRu$_2$Si$_2$.
Analysis of the resistivity data shows a pronounced decrease of the electronic quasiparticle mass across $H_c$.
These results establish CeTiGe as a new metamagnetic Kondo-lattice system, with an exceptionally large, metamagnetic
transition of first-order type at a moderate field.
\end{abstract}
\maketitle
\begin{figure}[t]
\begin{center}
\includegraphics[width=\columnwidth,angle=0]{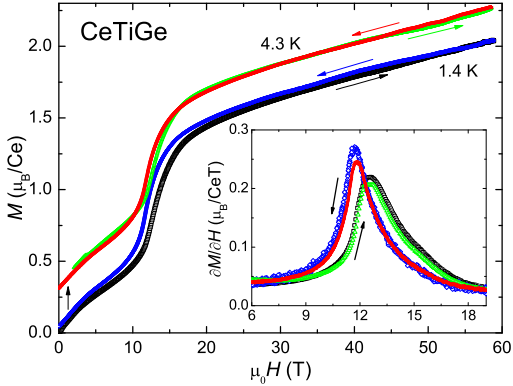}
\end{center}
\caption{(color online) Magnetization $M(H)$ of CeTiGe in fields up to $\mu_{0}H = 60$\,T measured at 1.4 and 4.3\,K.
The data at 4.3\,K were shifted upwards by 0.25\,$\mu_{B}$/Ce. The arrows indicate the field direction.
A step with $\Delta M$ = 0.74\,$\mu_B$/Ce is observed at a critical field $\mu_0 H_c$ = (12.5 $\pm$ 0.5)\,T.
The inset shows the derivative $\partial M/\partial H$ vs $H$ in the field range $6 < \mu_{0}H < 19$\,T for both temperatures.}
\label{fig1}
\end{figure}

Kondo lattices are intermetallic compounds based typically on a rare-earth element with an unstable \textit{f}-shell, usually
Ce or Yb. Because of this instability they can be tuned from a magnetic to a non-magnetic state by applying mechanical or
chemical pressure. At the crossover, one observes very unusual properties, e.g., the formation of heavy quasiparticles
(heavy fermions), unconventional superconductivity and non-Fermi-liquid ground states \cite{Loehneysen2007, Gegenwart2008}.
Accordingly, these systems have been the subject of continuous intensive research since more than 30 years.\\
Most recently, the fate of the 4\textit{f} electrons, i.e. how the related degrees of freedom are removed from the Fermi
surface through localization of the $f$-electrons, has developed into a central issue in the field \cite{Loehneysen2007,Gegenwart2008}. In this context,
another unusual phenomenon in Kondo lattices has recently regained considerable attention: The pronounced metamagnetic
transition (MMT) observed in the paramagnetic heavy-fermion (HF) system CeRu$_2$Si$_2$ \cite{Haen1987}. At a critical field
$H_c$ = 7.7 T applied along the tetragonal $c$-axis, which is the easy axis for this Ising-type system, one observes a large
step in the field-dependent magnetization $M(H)$ of the order of 0.4\,$\mu_{\rm B}$/Ce. Very accurate measurements down to
lowest temperatures indicate that this transition is not a real thermodynamic phase transition, but rather a sharp crossover,
since the transition width and height remain finite down to the lowest temperatures \cite{Sakakibara1995}. Despite intensive
studies over the last 20 years, the nature of this transition is still a matter of discussion. Strong changes of the Fermi
surface at $H_c$ were taken as evidence for a transition from an itinerant 4$f$ at $H < H_c$ to a localized 4\textit{f}
state at $H > H_c$ \cite{Aoki1993}. On the contrary, a recent study of the magnetoresistance suggests that this transition
corresponds to a continuous disappearance of one of the spin-split sheets of the Fermi surface \cite{Daou2006}.
One of the problems for a more general study of this type of MMT is the absence of further good examples in 4$f$-based Kondo
lattices at accessible fields. Magnetization curves with a sharp step in $M(H)$ have been reported for CeNi$_2$Ge$_2$
\cite{Fukuhara1996} and CeFe$_2$Ge$_2$ \cite{Sugawara1999}. However, the critical fields are much larger (42 and 30 T,
respectively), and the steps in $M(H)$ are significantly smaller ($ \leq 0.2$\,$\mu_{\rm B}$/Ce). Therefore, the MMTs in
these systems have not been deeply investigated. In literature many other Kondo lattices have been proposed to exhibit a MMT,
like, e.g., CeCu$_6$ \cite{Schroeder1992} or YbRh$_2$Si$_2$ \cite{Tokiwa2005}, but these compounds show only a kink in $M(H)$,
not a step. This means that some essential ingredients are missing in those systems, which are present in CeRu$_2$Si$_2$.\\
In this letter we report the discovery of a new example of a paramagnetic Kondo lattice which shows a pronounced metamagnetic
transition: CeTiGe. Our recent comprehensive study of the physical properties of CeTiGe established this compound as a new
paramagnetic HF system with a Sommerfeld coefficient $\gamma \approx 0.3$\,J/K$^2$mol~\cite{Deppe2009}, being located on
the non-magnetic side of the quantum critical point (QCP) within the Doniach phase diagram. A maximum in the susceptibility
$\chi(T)$ at 25\,K and in the 4\textit{f} specific heat $C_{4f}(T)/T$ at 16 K, as well as a single maximum in the
thermopower $S(T)$ at 17\,K (with a huge peak value of 60 $\mu$V/K) indicate a large degeneracy of the local moment involved.
Both $\chi(T)$ and $C_{4f}(T)$ could be well fitted with the Coqblin-Schrieffer model for $J = 5/2$ and $T_{0}$ = 82\,K,
corresponding to a Kondo scale $T_{\rm K} = 55$\,K~\cite{Deppe2009}. Preliminary results, indicating the presence of a MMT,
have been reported in Ref.~\cite{Deppe2010}. The new findings presented here are (i) a large step in the magnetization
$\Delta M(H) = 0.74 $\,$\mu_{\rm B}$/Ce at a quite low field of about 12.5\,T which is accessible with a standard
superconducting magnet, (ii) the MMT is accompained by a hysteresis in magnetization, magnetoresistance and magnetostriction
loops which strongly suggests the presence of a real thermodynamic first-order phase transition instead of the crossover
reported for CeRu$_2$Si$_2$~\cite{Flouquet2005}, (iii) the analysis of the resistivity data shows an abrupt decrease
of the electronic quasiparticle mass across the MMT. Hence our discovery opens a new door for the study of this unusual
phenomenon.\\
The preparation and characterization of the samples is described in detail in Ref.~\cite{Deppe2009}. The magnetoresistance
and the linear magnetostriction were measured between $0.02$ and 4\,K in a magnetic field up to 18\,T in a $^3$He/$^4$He
dilution refrigerator using a standard 4-terminal ac technique and a high-resolution CuBe capacitance dilatometer, respectively.
The magnetization experiments were performed on powdered samples in a pulsed 60\,T-magnet down to 1.4\,K at the Dresden
High Magnetic Field Laboratory~\cite{Zherlitsyn2006}. The pulsed field data were scaled by absolute values of the
magnetization taken in a SQUID magnetometer.\\
%
The essential result is displayed by the magnetization shown in Fig.~\ref{fig1} as $M(H)$ versus $H$ for $T = 1.4$ and 4.3\,K.
The prominent feature is a step-like anomaly at a critical field of $\mu_0 H_c$ $\approx$ 12\,T. At lower and higher
fields $M(H)$ increases weakly and almost linearly with $H$. The value at the highest field is slightly lower than the
saturation magnetization expected for a fully localized Ce$^{3+}$ (2.14 $\mu_{\rm B}$/Ce). Extrapolating the linear
$M(H)$ towards $H_c$ an idealized magnetization step size of $\Delta M$ = 0.74 $\mu_{\rm B}/{\rm Ce}$ at 1.4\,K can be
extracted. This is almost twice the value reported for CeRu$_2$Si$_2$, and a factor of four larger than in
CeNi$_2$Ge$_2$ \cite{Fukuhara1996} or in CeFe$_2$Ge$_2$ \cite{Sugawara1999}.
\begin{figure}[t]
\begin{center}
\includegraphics[width=\columnwidth,angle=0]{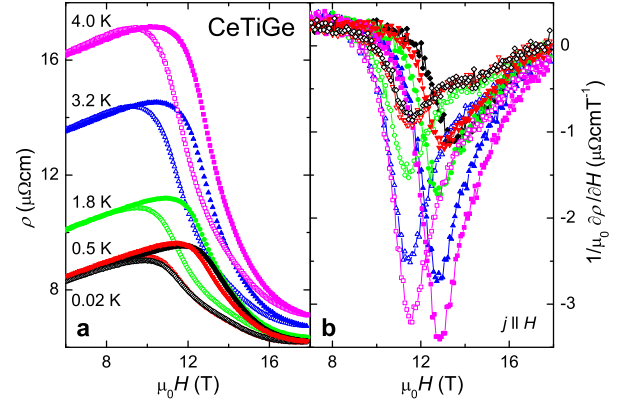}
\end{center}
\caption{(color online) \textbf{a)} Magnetoresistance $\rho(H)$ of CeTiGe polycrystals with current $j~\|~H$ measured at different
temperatures between 0.02 and 4\,K. A hysteresis loop between data collected with increasing field (closed symbols) and
those with decreasing field (open symbols) is observed for all $T$. \textbf{b)} The derivative $\partial \rho / \partial H$
of the magnetoresistance is plotted in the field range $7 \leq \mu_{0}H \leq 18$\,T.} 
\label{fig2}
\end{figure}
Magnetization data taken on increasing and decreasing magnetic field show a hysteresis loop.
This is better seen in the differential susceptibility $\partial M/\partial H$ displayed in the inset of Fig.~\ref{fig1}
for the field range $6 \leq \mu_{0}H \leq 19$\,T: A sharp peak at 12.5\,T for increasing field and one at 11.7\,T for
decreasing field yield a metamagnetic transition with a hysteresis width of $\Delta\mu_{0}H \approx$ 0.8\,T, inferring
a transition of the first-order type. The difference in peak height and width between increasing and decreasing $H$ is likely
related to the asymmetric field pulse which presents a steeper rising edge~\cite{Zherlitsyn2006}, and to magnetocaloric
effects. The peak positions are identical for $T$ = 1.4 and 4.3\,K, indicating a vertical phase boundary in the $H-T$ phase
diagram. Therefore, the hysteresis cannot result from a shift in the temperature during the measurement. The $M(H)$ values
at 4.3\,K are slightly lower than those at 1.4\,K, but the difference is within the absolute accuracy of the measurement.
A small splitting opens below 47\,T in the data at 1.4\,K, which might be related to the magnetocaloric effect or to the
alignment of the powder at high field, since the data for 1.4\,K were taken first.\\
The short time scale of pulsed field experiments induces non-equilibrium effects, e.g., temperature shifts due to the
magnetocaloric effect. For this reason we focused on experiments at very low temperatures and in static
magnetic fields. Fig.~\ref{fig2}a shows the magnetoresistance between 6 and 18\,T at temperatures in the range 
$0.02 \leq T \leq 4$\,K. In order to minimize the effects due to the Lorentz force we have probed the longitudinal
magnetoresistance with the current density $j~\|~ H$. In all measurements the field was ramped up and down with a slow
sweeping rate of 0.5\,mT/s. The most prominent feature in Fig.~\ref{fig2}a is a sharp drop of the resistivity at
$\mu_0 H_c$ $\approx$ 12\,T with a pronounced hysteresis opening between the up and down field sweeps.
The derivative $\partial \rho / \partial H$, plotted for the field range $7 \leq \mu_{0} H \leq 18$\,T in Fig.~\ref{fig2}b,
emphasizes the sharpness of the transition by showing striking peaks at 12.9\,T for increasing field and at 11.5\,T
for decreasing field, reminiscent of those observed in $M(H)$. Thus, the mean value of the critical field
$\mu_0 H_c$ = 12.2\,T is the same as in $M(H)$, while the hysteresis is even larger.
While the peak positions do not significantly shift with temperature, as for the magnetization, the peak size, i.e. the size
of the drop in $\rho(T)$, increases strongly with increasing $T$, in contrast to the behavior of $M(H)$. However,
this increase can be explained straightforwardly: It is directly connected with the strong increase of the zero field
resistivity with temperature, while the resistivity measured at fields larger than $H_c$ increases only very weakly with $T$.
This effect deserves a deeper analysis, since it gives further insight into the nature of the metamagnetic transition.
In Fig.~\ref{fig3}a the temperature dependence of the resistivity is plotted as $\rho(T)$ vs $T^2$ for selected fields
below and above $H_{c}$. All curves follow the $T^2$ power law expected for a Fermi liquid over two decades of temperature.
All plots for fields below the MMT show almost the same slope, being just slightly shifted upwards with
increasing field. In contrast, the slope of the plots for $H > H_c$ is much smaller.
\begin{figure}[t]
\begin{center}
\includegraphics[width=\columnwidth,angle=0]{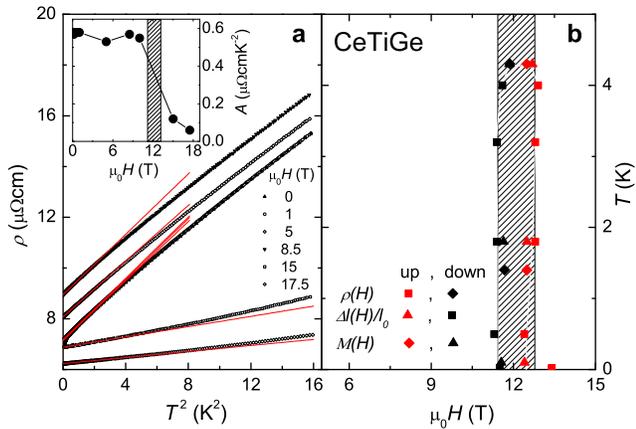}
\end{center}
\caption{(color online) \textbf{a)} Electrical resistivity $\rho(T)$ plotted against $T^2$ for different fields up to 17.5\,T.
Symbols indicate the experimental data, whereas lines represent the linear fits for $T \rightarrow 0$.
The inset shows the field dependence of the slope of the low-$T$ fits ($A$ coefficient). \textbf{b)} $H-T$ phase diagram of
CeTiGe below 4.3\,K deduced from $M(H)$, $\rho(H)$ and $\Delta l(H)/l_0$. The hatched area indicates the hysteresis region.} 
\label{fig3}
\end{figure}
In a Fermi liquid picture this slope
is related to the quasiparticle - quasiparticle scattering cross section, and it is proportional to the square of the
quasiparticle effective mass. In CeTiGe this relation was shown to be fulfilled at $H = 0$ \cite{Deppe2009}. Thus, the drop
from a large slope for $H < H_c$ to a small slope for $H > H_c$ indicates a pronounced drop of the effective mass of the
quasiparticles. For a quantitative analysis we fitted the low-$T$ part ($T < 1.5$\,K) with the function
$\rho(T) = \rho_{0} + A(H) \cdot T^2$ and plotted in the inset of Fig.~\ref{fig3}a the field dependence of the $A$ coefficient.
It drops from 0.6 to 0.06 $\mu\Omega$\,cmK$^{-2}$, which implies at least a decrease of the effective mass by a factor of 3.
Hence, the MMT is linked to a pronounced decrease of the quasiparticle density of states at the Fermi level. 
A further analysis demonstrates that the magnetoresistance at fields $H < H_c$ follows the behavior expected for a normal
metal with a Fermi liquid ground state. At all temperatures investigated here $\rho(H)$ first increases quadratically but
then evolves towards a linear $H$ dependence before $H$ approaches $H_{c}$. The region with quadratic $H$-dependence is
larger for higher $T$. This suggests a scaling behavior. We therefore plotted the data according to Kohler's rule:
$(\rho(H) - \rho(0))/\rho(0) = f(H/\rho(0))$ (not shown). All the data for T = 0.02, 0.5 and 1.8\,K fall onto the same curve,
almost up to $H_c$, while the data for T = 3.2 and 4\,K exhibit only a slight upward deviation.
Thus, for $H < H_c$, $\rho(H)$ follows perfectly Kohler's rule, confirming a Fermi liquid state and the absence of significant
spin disorder scattering.\\
In metallic systems metamagnetic transitions from a paramagnetic to a polarized state are usually connected with a large
magnetostriction \cite{Goto1998}. We therefore investigated the effect of the magnetic field on the length of a CeTiGe
sample. The linear magnetostriction $\Delta l(H)/l_0$ was measured on a CeTiGe polycrystal with a length of 4\,mm at
different temperatures between 0.1 and  4.3\,K. Afterwards, the sample was rotated by 90 degrees and the measurements were
repeated. The relative length changes $\Delta l(H)/l_0$ vs $H$ observed at 0.1\,K are plotted in Fig.~\ref{fig4}a.
\begin{figure}[t]
\begin{center}
\includegraphics[width=\columnwidth,angle=0]{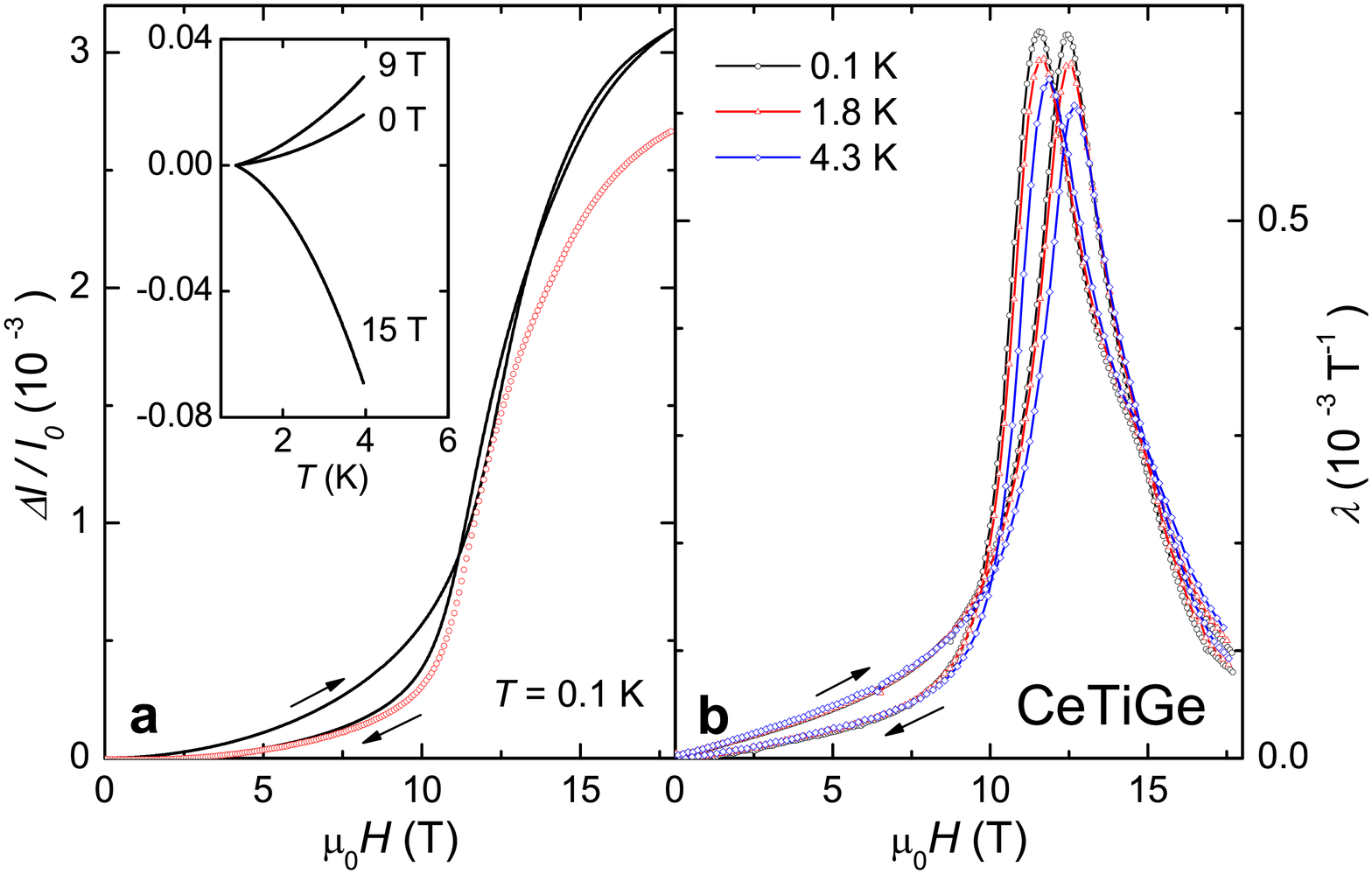}
\end{center}
\caption{(color online) \textbf{a)} Linear magnetostriction $\Delta l(H)/l_0$ of CeTiGe measured up to $\mu_0 H$ = 18\,T at
$T$ = 0.1\,K for two perpendicular sample orientations (black lines and red circles). Inset: The linear thermal expansion
$\Delta l(T)/l$ vs $T$ is plotted in the temperature range $1 < T < 4$\,K for three selected fields of 0, 9 and 15\,T:
$\Delta l(T)/l$ changes sign across the MMT. \textbf{b)} The magnetostriction coefficient $\lambda(H) = 1/\mu_0 l_0 \cdot \partial
l/ \partial H$ is plotted as a function of $H$  for the up and down sweeps at three selected temperatures: 0.1, 1.8 and
4.3\,K.} 
\label{fig4}
\end{figure}
The field dependence of the magnetostriction is completely dominated by a huge, step-like increase in the field range
between 10 and 15\,T. In contrast, below 10\,T the magnetostriction is comparatively small, while saturation seems to occur
beyond 18\,T. Thus, the MMT is connected with a step-like increase in $\Delta l(H)/l_0$ of the order of $2.0 \cdot 10^{-3}$,
which is almost one order of magnitude larger than the one observed in CeRu$_2$Si$_2$ \cite{Lacerda1989}. Rotating the sample
does not affect the overall behavior, only the magnitude gets slightly reduced, indicating that the polycrystals do not
have a significant texture. Furthermore, the curves for increasing and decreasing field differ slightly, showing a
hysteresis. This is better seen in the linear magnetostriction coefficient $\lambda(H) = 1/\mu_0 l_0 \cdot
\partial l/ \partial H$ shown in Fig.~\ref{fig3}b. It displays sharp peaks at $\mu_0 H_c \approx 12.3$ and 11.4\,T for
increasing and decreasing fields. The hysteresis width is about 0.9\,T. As in magnetization there is almost no change with
increasing temperatures up to 4.3\,K, except for a small decrease of the peak height.\\
According to magnetoelastic theory one expects the magnetostriction to be proportional to the square of the magnetization,
$1/\mu_0V \cdot \partial V/ \partial H = K \cdot M^2(H)$, where $K$ is the magnetoelastic coupling constant
\cite{Wohlfarth1969}. A preliminary analysis reveals such a quadratic dependence both at low and at high fields.
$K$ has a similar value below and above the transition, of the order of $5 \cdot 10^{-3}$ (Ce/$\mu_{\rm B})^2$, which is of
the same size as those reported for the 3\textit{d} metamagnetic systems YCo$_2$ and LuCo$_2$ \cite{Goto1998}.\\
We also performed preliminary measurements of the thermal expansion. In the inset of Fig.~\ref{fig4} we show the thermal
expansion $\Delta l(T)/l_0$ at constant fields of 0, 9 and 15\,T, i.e. well below, just below, and above the MMT, for
$0.05 < T < 4$\,K. The thermal expansion is positive at zero field, becomes larger with increasing field, but changes
sign and becomes strongly negative for $H > H_{c}$. The large positive thermal expansion at zero field is typical for
Ce-based Kondo-lattice systems. In a simple approach this can be related to the larger volume of the trivalent state compared
to that of an intermediate valent state, and to the slight shift of the valence towards 3+ with increasing temperature.
A pronounced change at the MMT from a large positive thermal expansion to a large negative one has already been observed
in CeRu$_2$Si$_2$ and is also, in a simple approach, related to the field induced stabilization of a "localized" trivalent
state. An analysis of the thermal expansion coefficient $\alpha /T$ = $1/l_0 \cdot \partial l/ \partial T$ using a plot
$\alpha(T)/T$ vs $T^2$ (not shown) shows the ratio $\alpha(T)/T$ to be almost independent of $T$, i.e. the thermal
expansion is completely dominated by the linear-in-$T$ electronic contribution expected in a Fermi liquid.
The ratio $\alpha(T)/T$ increases with $H$, from $2 \cdot 10^{-6}$\,K$^{-2}$ at $H$ = 0 up to $4.6 \cdot 10^{-6}$\,K$^{-2}$
for fields just below $H_{c}$, but then switches to a large negative value of about $-11 \cdot 10^{-6}$\,K$^{-2}$ at fields
just above $H_{c}$. Further increasing the field leads to a reduction of the absolute value, with $\alpha(T)/T = -3.3 \cdot
10^{-6}$ K$^{-2}$ at $\mu_{0}H = 18$\,T. The weak $T$ dependence of $\alpha(T)/T$ below 4\,K in CeTiGe, even for fields close
to the critical field, contrasts the large $T$ dependence reported for CeRu$_2$Si$_2$ close to $H_c$ in the same $T$ range
\cite{Paulsen1990,Weickert2010}. This likely reflects the difference between a continuous crossover in CeRu$_2$Si$_2$,
which implies strong $T$ dependent fluctuations near the critical field, and the first-order transition in CeTiGe,
for which no fluctuations are expected.\\
An interesting parameter in Kondo lattices is the Gr\"uneisen ratio $\Gamma = k \cdot V_{mol} \cdot \beta(T)/C(T)$,
where $k$ is the compressibility, $V_{mol}$ the molar volume and $\beta(T)$ the volume thermal expansion coefficient.
This ratio between thermal expansion and specific heat reflects the pressure dependence of the characteristic energy
of the system \cite{deVisser1990}. With $\beta(T)/T=3\cdot\alpha(T)/T=6\cdot10^{-6}$\,K$^{-2}$, $\gamma=0.3$\,J/K$^2$mol,
$V_{mol} = 41 \cdot 10^{-6}$\,m$^3$/mol, and assuming $k=100$\,GPa (a typical value for intermetallic Ce-based systems)
we obtain a Gr\"uneisen ratio $\Gamma \approx 85$ at zero field. Such a large $\Gamma$ is typical for Kondo lattices close
to the transition between a magnetic and a paramagnetic ground state \cite{deVisser1990}. However, it is a factor of 2
smaller than in CeRu$_2$Si$_2$, likely reflecting the absence of critical behavior in CeTiGe.\\
In Fig.~\ref{fig3}b we show the preliminary $H-T$ phase diagram as deduced from the magnetization, magnetoresistance and
magnetostriction results. The critical fields obtained from the peak positions in the $H$-derivative of these properties
agree nicely and evidence a vertical phase boundary. Furthermore, there is no significant increase in the transition width
or the hysteresis width with temperature at least up to 4.3\,K, in contrast to the significant broadening observed
in CeRu$_2$Si$_2$ in the same $T$ range. This is a further evidence for the presence of a true thermodynamic phase transition
in CeTiGe.\\
In summary, our investigation of the paramagnetic heavy-fermion system CeTiGe reveals a pronounced metamagnetic transition
at a moderate field $\mu_{0}H_{c} = 12.5$\,T. At this field we observe a large increase in the magnetization
$\Delta M = 0.74$\,$\mu_{B}$/Ce and in the magnetostriction $\Delta l/l_{0} = 2.0 \cdot 10^{-3}$, which are even larger
than in the prototypical heavy-fermion metamagnet CeRu$_2$Si$_2$, as well as a pronounced drop in the magnetoresistance.
The opening of a hysteresis between increasing and decreasing field data in all investigated properties indicate this MMT
to be a real thermodynamic transition of the first-order type, in contrast to the crossover behavior reported for
CeRu$_2$Si$_2$. Increasing the temperature from 0.02 to 4.3\,K leaves the transition almost unaffected: $H_{c}$ is nearly
independent of $T$ and neither the hysteresis width nor the transition width increases with $T$, which is incompatible with
crossover behavior. In addition, the thermal expansion in fields close to $H_{c}$ do not show evidence of critical fluctuations
up to 4.2 K, giving further support to the first-order type of the transition. The analysis of the resistivity and
magnetoresistance data points to a pronounced drop of the effective mass of the electronic quasiparticles across the
transition, while thermal expansion and magnetostriction suggest that this drop might be connected with some kind of
localization of the 4\textit{f} electrons. The large Gr\"uneisen parameter $\Gamma \approx 85$, deduced from the thermal
expansion and specific heat data, is in accordance with CeTiGe being close to the transition from a paramagnetic to
a magnetically ordered ground state. The first-order nature of the metamagnetic transition as well as the exceptionally
large anomalies in $M(H)$ and $l(H)$ establish CeTiGe as a rather unique metamagnetic system among the 4\textit{f}-based
Kondo-lattice compounds.\\
The authors thank U. Burkhardt and P. Scheppan for detailed microprobe studies. FW acknowledges finacial founding by the
MPG Research Initiative \textit{Materials Science and Condensed Matter Research at the
Hochfeldmagnetlabor Dresden}. Part of this work has been supported by
EuroMagNET II under the EC contract 228043 and by the DFG Research Unit 960 ``Quantum Phase Transitions''.
\bibliographystyle{h-physrev}
\bibliography{CeTiGe_PRL}
\end{document}